\documentstyle[aps,pre,twocolumn,epsf]{revtex}

\newcommand \be {\begin{equation}}
\newcommand \ee {\end{equation}}
\newcommand \ben {\begin{eqnarray}}
\newcommand \een {\end{eqnarray}}
\newcommand \nline {\nonumber \\}

\newcommand \bee {\delta a}
\newcommand \triangular {hexagonal }
\newcommand \psitee {\psi_h}
\newcommand \qtee {q_h}
\newcommand \Atee {A_h}

\begin{document}
\draft
\twocolumn[\hsize\textwidth\columnwidth\hsize\csname
@twocolumnfalse\endcsname

\title{Modeling Elasticity in Crystal Growth}

\author{K. R. Elder$^1$, Mark Katakowski$^1$, Mikko Haataja$^2$ 
and Martin Grant$^2$.}

\address{$^1$Department of Physics, Oakland University, Rochester, MI,
48309-4487}
\address{$^2$Physics Department, Rutherford Building, 3600 rue
University, McGill University,
Montr\'eal, Qu\'ebec, Canada H3A 2T8.}

\date{\today}
\maketitle

\begin{abstract}
A new model of crystal growth is presented that describes 
the phenomena on atomic length and diffusive time scales.  
The former incorporates elastic and plastic deformation 
in a natural manner, and the latter enables access
to times scales much larger than conventional atomic 
methods.  The model is shown to be consistent with the 
predictions of Read and Shockley for grain boundary energy, 
and Matthews and Blakeslee for misfit dislocations in 
epitaxial growth.
\end{abstract}


\vskip1pc]


	The appearance and growth of crystal phases occurs in many 
technologically important processes including epitaxial growth 
and zone refinement.  While a plethora of models have been constructed 
to examine various aspects of these phenomena it has proved difficult to
develop a computationally efficient model that can be used for a
wide range of applications.  For example, standard molecular dynamics
simulations include the necessary physics but are limited by 
atomic sizes (\AA) and phonon time scales (ps). 
Conversely continuum field theories can access longer length (i.e., 
correlation length) and time (i.e., diffusive) scales, but are difficult 
to incorporate with the appropriate physics.   In this paper a new model 
is presented that includes the essential physics and is 
not limited by atomic time scales.  

To illustrate the features that must be incorporated it is useful to 
consider two examples.  First consider the nucleation and growth of 
crystals from a pure supercooled liquid or vapor phase.	In such a process
small crystallites nucleate (heterogeneously or homogeneously) and grow
in arbitrary locations and orientations.  
Eventually the crystallites impinge on one another
and grain boundaries are formed.  Further growth is then determined
by the evolution of grain boundaries.  Now consider the growth of a 
thin crystal film on a substrate of a different crystal structure. 
Typically the substrate stresses the over-lying film which can 
destabilize the growing film and cause an 
elastic defect-free morphological deformation \cite{at72,GRINFELD}, 
plastic deformation involving misfit dislocations\cite{mb74}, or
a combination of both.  Thus 
the model must be able to nucleate crystallites at arbitrary 
locations and orientations and contain elastic and plastic deformations.
While all these features are naturally incorporated in 
atomistic descriptions,
they are much more difficult to include in continuum or phase field models.

	Historically, many continuum models have been developed
to describe certain aspects of crystal growth and liquid/solid
transitions in general.	 At the simplest level, `model A' in the
Halperin and Hohenberg\cite{hh77} classification scheme has been
used to describe liquid/solid transitions.  This model treats all solids 
equivalently and does not introduce any crystal anisotropy.  
Extensions to this basic model have been developed to incorporate a 
solid phase that has multiple states that represent 
multiple orientations\cite{mesg95,cy94} or recently\cite{wck98} 
an infinite number of orientations.  Unfortunately these models 
do not properly include plastic and elastic deformations.  
Other models\cite{mg99,ossr99,j00,ssla01,wjck01} 
have sought to include elastic and plastic deformations with reference 
to a specific reference lattice, but cannot account for multiple orientations.  

\begin{figure}[btp]
\epsfxsize=8cm \epsfysize=8cm
\epsfbox{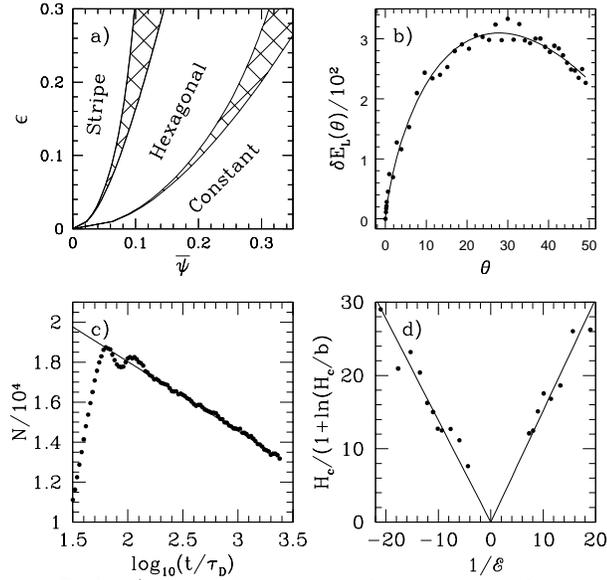}
\caption{a) Mean field phase diagram.
In this figure the `hatched' regions are 
coexistence regions. b) Grain boundary energy. 
The points are from numerical simulations and the line a 
fit to the Read-Shockley equation.  c) Grain growth. 
In this figure the number defects is plotted and 
the solid line is a guide to the eye.
d) Epitaxial growth. The points 
are from numerical simulations and the lines are 
best fits.}  
\label{fig:evth}
\end{figure}

	In this work these limitation are overcome by considering a free 
energy that is minimized by a periodic \triangular (i.e., solid) state.  
Such free energies have arisen in many other physical systems\cite{hex,sd95} 
(such as water/surfactant systems, copolymers, Rayleigh-B\'enard convection and 
ferromagntic films) and in some instances have even been described in 
terms of 
crystalline terminology\cite{sd95}.  The model used in this work describes 
the statics and dynamics of a conserved field, $\psi$, by the following free energy 
and equation of motion,
\be
\label{eq:free}
 {\cal F} = \int d\vec{r} \left(\psi\left[(q_o^2+\nabla^2)^2-\epsilon
\right]\psi/2+\psi^4/4\right)
\ee
and
\be
\label{eq:eom}
\partial \psi/\partial t = 
\nabla^2\left(\delta {\cal F}/\delta \psi\right) + \eta,
\ee
where $\eta$ is a stochastic noise, with zero mean and 
correlations $<\eta(\vec{r},t)\eta(\vec{r}',t')> = 
- G \nabla^2 \delta(\vec{r}-\vec{r}')\delta(t-t')$,
and $G=0$ hereafter, $q_o$ and $\epsilon$ are constants. 
The field $\psi$ represents the average atomic positions.
The focus of this paper is on two dimensions (2D);
it is straightforward to extend these calculations to 3D.  
In two dimensions, this
free energy is minimized by striped ($\psi_s$), 
\triangular ($\psitee$) and 
constant ($\psi_c$) states depending on the average value, $\bar{\psi}$, of 
$\psi$.  To estimate the phase diagram these states can be approximated 
$\psi_{s} = A_s \sin(q_s x) + \bar{\psi}$
\be
\label{eq:hex}
\psi_{h} = \Atee \left[\cos\left(\qtee x\right)
\cos\left(\frac{\qtee y}{\sqrt{3}}\right)
+\frac{1}{2}\cos\left(\frac{2\qtee y}{\sqrt{3}}\right)\right]+\bar{\psi}
\ee 
and $\psi_c=\bar{\psi}$.  
Substituting these expressions into 
Eq. (\ref{eq:free}) and minimizing subject to the constraint 
$\int d\vec r \psi = \bar{\psi}$ 
gives 
the values for the characteristic amplitudes $A$ and wavenumber
$q$\cite{minz} and the phase diagram show in 
Fig.\ (\ref{fig:evth}a).

	The linear elastic properties of the \triangular phase can be 
determined by calculating the change of energy under shear, bulk or 
dilational deformations within the two mode approximation given in 
Eq. (\ref{eq:hex}).  
It is straightforward to find the elastic moduli for the isotropic
solid:
$C_{12} = C_{44} = C_{11}/3$, where
\be
C_{12} = \left[\left(3\bar{\psi}
+\sqrt{15\epsilon-36\bar{\psi}^2}\right)q_o^2\right]^2/75.
\ee
For these coefficients\cite{CHAIKIN} the Poisson ratio 
is $\nu = 1/3$ and the shear modulus is $\mu = C_{12}$.  

	The energy per unit surface length, $E_L$, between grains that 
differ in orientation by an angle $\theta$ was determined numerically 
and, in Fig.\ (\ref{fig:evth}b), compared with the prediction of
Read and Shockley\cite{rs50}, i.e., $E_L = E_M \theta[1-\ln(\theta/\theta_M)]$ 
where $E_M$ and $\theta_M$ are constants.
The parameters of the simulations were 
$(\epsilon,\bar{\psi},q_o) = (4/15,1/4,1)$. In all the simulations conducted 
in the paper the time and space size were $\Delta t = 0.01$ and 
$\Delta x = \pi/4$ respectively.  
The Read-Shockley form
closely fits the data for $\theta_M = 27.85^o$ and $E_M = 0.064$.
The maximum angle, $\theta_M$ is similar to those observed in 
experiment\cite{rs50}.  The maximum energy/length, $E_M$, 
can be estimated using the 
Read-Shockley equation\cite{rs50} in 2D\cite{CHAIKIN}
using the elastic constants estimated above, 
$E_M = \mu(1+\nu)b/4\pi = C_{12}b/3\pi$,
where $b = 2\pi /\qtee$ is the magnitude of the Burger's vector.
For the parameters used in the simulations 
this gives $E_M=0.044$, as compared to our measured value of 0.064.

	The main advantage of the current approach over molecular dynamics 
simulations is the time scales that are accessible.   To illustrate 
this point it is useful to calculate the diffusion time using a 
standard Block-Floqu\'et linear analysis.  In such an approach the dynamics of a 
perturbation ($\delta \psi$) around an equilibrium crystal state ($\psi_{eq}$) 
is obtained by first substituting 
$\psi = \delta \psi + \psi_{eq}$ into Eq. (\ref{eq:eom}) and linearizing 
in $\delta\psi$.  
Substituting appropriate forms for the equilibrium state 
(i.e., $\psi_{eq} = \bar{\psi} + \sum a_{n,m} {\rm exp}[i(nq_x x+m q_y y)]$) 
and perturbation along one of the three principle axes 
(i.e., $\delta \psi = \sum 
\bee _{n,m}(t) {\rm exp}[i(n(q_x+Q) x+mq_yy]$)
and integrating over ${\rm exp}[i(k q_x x + l q_y y)]$ gives an 
equation of motion for the modes $\bee _{n,m}$.
Using the approximation for $\psi_{eq}$ given in Eq. (\ref{eq:hex}) 
and four modes (i.e., $(\bee _{1,1}$, $\bee _{-1,1}$, $\bee _{1,-1}$,
$\bee _{-1,-1}$) to describe 
the perturbation leads to a set of four linear ordinary differential equations that 
can be solved analytically, assuming $\bee _{n,m} \sim {\rm exp}(\omega t)$.
The largest eigenvalue is equal to $\omega = -3 q_o^4 Q^2$ 
implying a diffusion constant of $3q_o^4$.  In terms of times steps, this implies 
it takes roughly 1000 times steps for a diffusion time, $\tau_D$\cite{diff}, for the 
simulation parameters used in this paper. 

\begin{figure}[btp]
\epsfxsize=8cm \epsfysize=8cm
\epsfbox{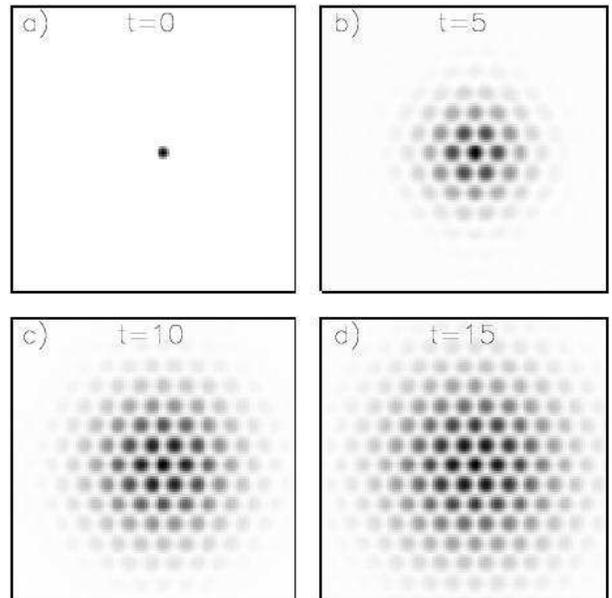}
\caption{Vacancy diffusion.  In this plot the gray scale is proportional to the 
difference between a perfect lattice and the one containing the vacancy at 
times $t/\tau_D=0, 5, 10$ and $15$ in a), b), c), and 
d) respectively. For clarity the magnitude of the gray 
scale was maximized in each figure.}
\label{fig:diffusion}
\end{figure}

	To confirm these calculations, vacancy diffusion was studied numerically 
by removing a single `particle' from a perfect \triangular lattice, for the 
parameters used in the grain boundary simulations.  
The dynamics of a vacancy is depicted in 
Fig.\ (\ref{fig:diffusion}).  It is important to note that this model
describes long times scales so the vacancy does not `jump' from site to site, 
but the probability of finding
the vacancy diffuses into the background matrix as 
shown in 
Fig.\ (\ref{fig:diffusion}).  The rate of spread can be used 
to numerically determine the diffusion constant and was $D=2.96$ 
which is quite close to the approximate calculation (i.e., $D=3$, for $q_o=1$). 

\begin{figure}[btp]
\epsfxsize=8cm \epsfysize=8cm
\epsfbox{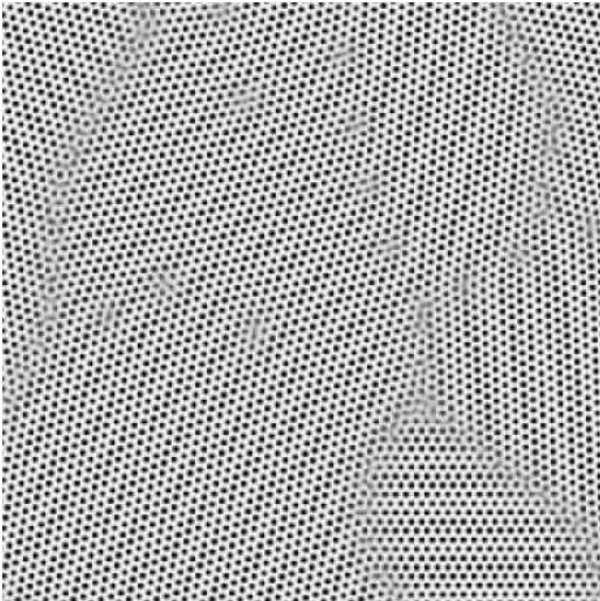}
\caption{Grain Growth. Gray scale plot of the order parameter $\psi$ 
at $t/\tau_D = 700$.  This plot contains one sixty-fourth of the 
full simulations cell.}
\label{fig:grain}
\end{figure}

	The ability of the continuum model to describe multiple 
crystals in arbitrary orientations and locations with the appropriate grain 
boundaries energies on diffusive time scales makes it ideal for the 
study of grain boundary growth.  To study this phenomena a system   
of size $4096 \Delta x \times 4096 \Delta x$ was simulated with parameters 
$(\epsilon,\bar{\psi},q_o)=(0.1,0.05,1)$. 
To begin the simulations, 256 crystals were nucleated at arbitrary 
locations by placing large fluctuations in $\psi$ at each nucleation site. 
As time evolves the small crystallites grow from the initial 
seeds until impingement.  Eventually the entire systems is filled with 
the \triangular phase and further evolution continues by motion at grain 
boundaries.  At this stage in the simulations there were approximately 
236,000 particles. To study the subsequent dynamics the number defects, 
where a defect is defined as a particle that does not have six nearest 
neighbors, was monitored.  The number of defects is shown in 
Fig.\ (\ref{fig:evth}c) 
as a function of the logarithm of time.  Initially ($\log(t)< 1.7$) 
the number of defect increases as the total droplet surface area 
increases.  When the droplets impinge, 
the number of defects begins to 
decreases as local rearrangements take place.  At later times 
($\log(t/\tau_D)>2.0$) 
the grain boundaries evolve at a very slow rate.   
In these simulations it is found that the number of defects decreases 
logarithmically at late times.

	The inclusion of elastic and plastic deformations 
allow the study of morphological instabilities in epitaxial growth. 
When a film grows 
on a bulk material that has a different lattice constant the film 
can become corrugated (``buckle'')
in an attempt to relieve the strain\cite{at72}.
The buckling of the surface relieves the strain in some regions but increases the strain 
in others.  At these locations dislocations eventually nucleate.  The 
critical height, $H_c$,  at which these nucleate 
is well described by the Matthews and Blakeslee equation\cite{mb74} which has the 
functional form, $H_c = H_o[1+\ln(H_c/b)]/{\cal E}$,
where $H_o$ is a constant and  ${\cal E}$ is the mismatch between the film and 
bulk lattice parameters, 
i.e., ${\cal E} = (a_{film}-a_{bulk})/a_{bulk}$, where $a$ is a lattice constant.

\begin{figure}[btp]
\epsfxsize=8cm \epsfysize=8cm
\epsfbox{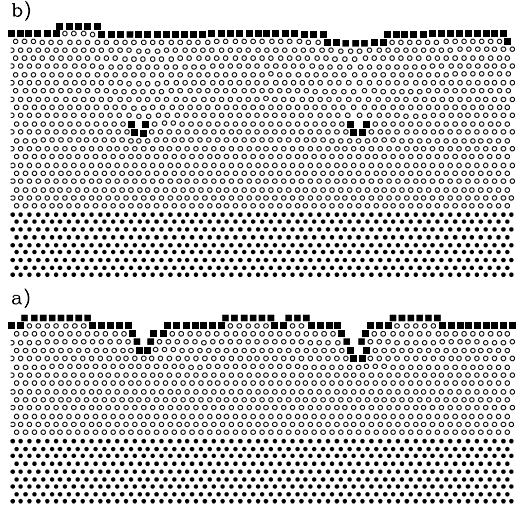}
\caption{Epitaxial Growth.  In this figure the 
film and bulk particles (defined as local maxima of $\psi$) are 
plotted as open and closed circles respectively.  Defects are 
plotted as solid squares.}
\label{fig:epitaxial}
\end{figure}

To study this phenomena a thin film with $q_o=q_f$ was grown on a bulk 
sample ($q_o=1$) with the parameters $(\epsilon,\bar{\psi})=(1/4,1/4)$ for 
various values of $q_f$.  The system had a width of $8192 \Delta x$ and 
height $1024 \Delta x$.  A small portion of a sample simulation 
is shown in Fig.\ 
(\ref{fig:epitaxial}) for a lattice mismatch 
of 6.4\% , 
(i.e., ${\cal E} = (2 \pi/q_f-2 \pi/q_o)/(2 \pi/q_o) = 0.064$).  
The buckling phenomena is shown in Fig.\ (\ref{fig:epitaxial}a) and the 
nucleation of dislocations in 
Fig.\ (\ref{fig:epitaxial}b).  
The critical height was defined as the height at which the interface 
velocity changes (i.e., since the number of particles arriving at 
the surface is conserved the velocity of the film interface changes 
when dislocations appear and change the total number of interface particles).   
The numerical results for $H_c$ shown in 
Fig.\ (\ref{fig:evth}d) are consistent with  
the functional relationship proposed by Matthews and Blakeslee\cite{mb74}.

\begin{figure}[btp]
\epsfxsize=8cm \epsfysize=8cm
\epsfbox{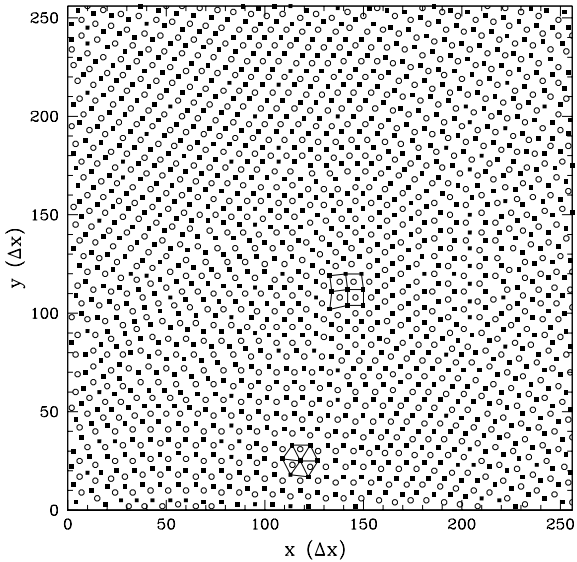}
\caption{Binary Alloy.  In this figure the open circles and solid squares are the maxima of 
$\psi_1$ and $\psi_2$ respectively.  Two different crystal structures have been 
highlighted by joining nearest neighbors of the same species.}
\label{fig:binall}
\end{figure}

	The one component model described by Eqs. (\ref{eq:free}) 
and (\ref{eq:eom}) does not support other metastable 
crystal phases and thus cannot be used to, for example, study 
structural phase transitions.  However it is 
straightforward to extend the model to include more than one kind 
of particle which can give rise to other metastable crystal phases. 
For example a binary system can be easily modeled by a free energy of the form;
\ben
 {\cal F} &=& \int d\vec{r} \left(\psi_1\left[(q_1^2+\nabla^2)^2-\epsilon_1
\right]\psi_1/2+{\psi_1^4}/{4}\right. \nline
&\,&+\left.\psi_2\left[(q_2^2+\nabla^2)^2-\epsilon_2
\right]{\psi_2}/{2}+{\psi_2^4}/{4}+\alpha \psi_1\psi_2\right) 
\een
and the equations of motion;
$\partial \psi_1/\partial t = \Gamma_1 \nabla^2 
\delta {\cal F}/\delta \psi_1+\eta_1$ and 
$\partial \psi_2/\partial t = \Gamma_2 \nabla^2 
\delta {\cal F}/\delta \psi_2+\eta_2$, 
where $\alpha$ is a coupling constant.  The properties (i.e., lattice 
constant, bulk compressiblity, etc.) of the individual atoms are 
controlled by the parameters with subscripts $1$ or $2$ and by the 
average value of $\psi_1$ and $\psi_2$.  When an individual binary alloy 
droplet is grown a hexagonal pattern typically emerges.  However when 
a random initial condition is used the patterns usually contain more than 
one crystal phase since the energy of the hexagonal state is very close 
to a face centered cubic.  One such a configuration is shown in Fig.\ 
(\ref{fig:binall}).  Eventually the system will evolve to 
a hexagonal state.  Thus the model can be used to study structural phase 
transitions.  

In conclusion, the preceding simulations and calculations have provided 
ample evidence that the model described by Eqs.\ (\ref{eq:free}) 
and (\ref{eq:eom}) provide an adequate description of crystal behavior 
on long time and atomic length scales.  Thus this model should provide 
a useful method of studying the crystallization of pure and multicomponent 
materials. 

This work was supported by a Research Corporation grant CC4787 (KRE),
and a NSF-DMR Grant 0076054 (KRE), the Academy of Finland (MH), the 
Natural Sciences and Engineering Research Council of Canada (MG) and 
{\it le Fonds pour la Formation de Chercheurs et l'Aide \`a la 
Recherche du Qu\'ebec\/} (MG),


\begin{references}

\bibitem{at72} R. J. Asaro and W. A. Tiller, Metall. Trans. {\bf 3}, 
1789 (1972).

\bibitem{GRINFELD}
M. Grinfeld, Dokl. Akad. Nauk. SSSR {\bf {265}},
836 (1982); Europhys.\ Lett.\ {\bf {22}}, 723 (1993).

\bibitem{mb74} J. W. Matthews and A. E. Blakeslee, 
J. Cryst. Growth {\bf 27}, 118 (1974).
J. W. Matthews, J. Vac.\ Sci.\ Technol.\ {\bf 12} 126 (1975).

\bibitem{hh77}
P. C. Hohenberg and B. I. Halperin, Rev. Mod. Phys. {\bf 49}, 435 (1977).

\bibitem{mesg95} B. Morin, K. R. Elder, M. Sutton and M. Grant, Phys.
  Rev. Lett. {\bf 75}, 2156 (1995).

\bibitem{cy94} L-Q. Chen and Wang, Phys. Rev. B {\bf 50}, 15752 (1994).

\bibitem{wck98} J. A. Warren, W. C. Carter, and R. Kobayashi,
Physica A {\bf 261}, 159 (1998); J. A. Warren, R. Kobayashi and W. C. Carter,
J. Cryst. Growth {\bf 211}, 18 (2000); R. Kobayashi, J. A. Warren and W. C. Carter,
Physica D {\bf 140}, 141 (2000).

\bibitem{mg99} J. M\"uller and M. Grant, 
Phys. Rev. Lett. {\bf 82}, 1736 (1999).
M. Haataja, J. M\"uller, A. D. Rutenberg, and M. Grant (preprint).

\bibitem{ossr99} D. Orlikowski, C. Sagui, A. Somoza, and C. Roland,
Phys. Rev. B {\bf 59}, 8646 (1999).

\bibitem{j00} A. E. Jacobs, Phys. Rev. B {\bf 61}, 6587 (2000).

\bibitem{ssla01} M. Sanati, A.Saxena, T. Lookman, and R. C. Albers, 
Phys. Rev. B {\bf 63}, 224114 (2001).

\bibitem{wjck01} Y. U. Wang, Y. M. Jin, A. M. Cuiti\~no, and A. G. Khachaturyan,
App. Phys. Lett. {\bf 78}, 2324 (2001).

\bibitem{hex} A. Linhananta and D. E. Sullivan, Phys. Rev. E {\bf 57}, 4547 (1998);
I. I. Potemkin and S. V. Panyukov, Phys. Rev. E {\bf 57}, 6902 (1998); J. Swift 
and P. C. Hohenberg, Phys. Rev. A {\bf 15}, 319 (1977).

\bibitem{sd95} C. Sagui and R. C. Desai, Phys. Rev. E {\bf 52}, 2807 (1995)

\bibitem{minz} In this approximation the free energy is 
minimized when $A_s = 2\sqrt{3\epsilon-9\bar{\psi}^2}/3$, 
$q_s=q_o$,  $\Atee = -4(b+\sqrt{15\epsilon-36\bar{\psi}^2}/3)/5$ 
and $\qtee = \sqrt{3}q_o/2$.

\bibitem{diff} In this paper a diffusion time, $\tau_D$ is 
defined as the time for a particle to diffuse one lattice 
constant and is equal to $a^2/D$ where $a$ is the lattice 
constant, i.e., $a \approx 2\pi/(\sqrt{3}q_o/2)$.

\bibitem{rs50}
W. T. Read and W. Shockley, Phys. Rev. {\bf 78}, 275 (1950).

\bibitem{CHAIKIN}
See, for example: {\it Principles of condensed matter physics\/},
P. M. Chaikin and T. C.  Lubensky (Cambridge University Press, Cambridge,
1995); {\it Theory of elasticity, 3rd edition\/} and L. D. Landau and
E. M. Lifshitz (Butterworth-Heinemann, Oxford, 1998). 

\end{references}
\end{document}